\documentclass[12pt]{article}
\begin{document}
\begin{titlepage}

\title{Difficulties of Teleparallel Theories of Gravity with Local Lorentz 
Symmetry}

\author{J. W. Maluf$\,^{(1)}$, S. C. Ulhoa$\,^{(2)}$, 
J. F. da Rocha-Neto$\,^{(3)}$ \\
and F. L. Carneiro$\,^{(4)}$\\
Instituto de F\'{\i}sica, 
Universidade de Bras\'{\i}lia\\
70.919-970 Bras\'{\i}lia DF, Brazil\\}
\maketitle
\bigskip
\bigskip

\begin{abstract}
A brief discussion is made about the relevance of surface terms in the 
Lagrangian and Hamiltonian formulations of theories of gravity. These surface 
terms play an important role in the variation of the action integral and in the
definition of field quantities such as the gravitational energy-momentum. Then 
we point out several inconsistencies of a recently proposed formulation of 
teleparallel theories of gravity with local Lorentz symmetry.

\end{abstract}
\thispagestyle{empty}
\bigskip
\vfill
\begin{footnotesize}
\par
\end{footnotesize}

\bigskip
{\footnotesize
\noindent (1) wadih@unb.br, jwmaluf@gmail.com\par
\noindent (2) sc.ulhoa@gmail.com\par
\noindent (3) rocha@fis.unb.br\par
\noindent (4) fernandolessa45@gmail.com}

\end{titlepage}
\newpage
\section{Introduction}
The search for alternative descriptions of Einstein's general relativity was 
initiated right after the advent of general relativity, with the 
proposal of the Kaluza-Klein theory. Teleparallel theories were 
considered by Einstein in 1928 as a possible geometrical 
set up for the unification of the electromagnetic and gravitational 
fields. Nowadays, extended or alternative formulations of general relativity
are investigated with the purpose of solving
cosmological problems, or establishing a possible quantum theory of
gravity, or even to address unsolved issues of the standard formulation of 
general relativity. The teleparallel equivalent of general relativity is one
such formulation which, among other features, allows the definitions of new 
geometrical quantities, constructed out of the torsion tensor, and provides a 
framework for defining energy, momentum and angular momentum of the 
gravitational field. In this article we address the issue of covariance of the
teleparallel equivalent of general relativity (TEGR) under Lorentz 
transformations. In order to analyse this issue, we briefly review in section 3
the importance of surface terms in
the action integral of gravitational theories. The discussion in section 3 will 
clarify the analysis of the problems regarding a surface term that appears in a 
recently proposed teleparallel theory with local Lorentz invariance.

\section{The TEGR}
The TEGR is a geometrical
formulation of the relativistic theory of gravity in terms of tetrad fields, 
which allows the notion of distant parallelism of vector or tensor fields.
The theory has a long history, which includes contributions from Hayashi 
and Shirafuji \cite{Hayashi}, Hehl \cite{Hehl}, Cho \cite{Cho} and 
Nester \cite{Nester}. See Ref. \cite{Maluf1} for additional relevant 
references.
The theory is defined by the field equations, which are equivalent to the 
Einstein's field equations of the metric formulation of general relativity. 
The set of tetrad fields is interpreted as a frame adapted to 
observers in space-time, and allows the projection of 
vector or tensors on the frame of an observer. The projection of a vector 
field $V^\mu(x^\alpha)$ on a certain frame, in the tangent space at 
the position $x^\alpha$, is given by 
$V^a(x^\alpha)=e^a\,_\mu(x^\alpha)\,V^\mu(x^\alpha)$. This is one of the main
geometrical properties of the set of tetrad fields.

The equivalence of the TEGR with the metric formulation of general relativity 
is based on a geometrical identity between the scalar curvature $R(e)$, 
constructed out of the tetrad fields, and an invariant combination of quadratic
terms in the torsion tensor, given by

\begin{equation}
eR(e) \equiv -e\left({1\over 4}T^{abc}T_{abc} + 
{1\over 2}T^{abc}T_{bac} - T^{a}T_{a}\right)
+ 2\partial_{\mu}(eT^{\mu})\,,
\label{1}
\end{equation}
where $ T_{a} = T^{b}\,_{ba}$, 
$T_{abc} = e_{b}\,^{\mu}e_{c}\,^{\nu}T_{a\mu\nu}$ and 
$T_{a\mu\nu}=\partial_\mu e_{a\nu}-\partial_\nu e_{a\mu}$. The latter is the 
torsion tensor, which is the anti-symmetric part of the Weitzenb{\"o}ck
connection $\Gamma^\lambda_{\mu\nu}=e^{a\lambda}\partial_\mu e_{a\nu}$, i.e.,
$T_{a\mu\nu}=e_{a\lambda}T^\lambda\,_{\mu\nu}=
e_{a\lambda}(\Gamma^\lambda_{\mu\nu}-\Gamma^\lambda_{\nu\mu})$.\par
\bigskip
\noindent (Notation: 
space-time indices $\mu, \nu, ...$ and SO(3,1) (Lorentz) indices $a, b, ...$ 
run from 0 to 3. The flat space-time metric tensor raises and lowers tetrad 
indices, and is fixed by $\eta_{ab}= e_{a\mu} e_{b\nu}g^{\mu\nu}=(-1,+1,+1,+1)$.
The frame components are given by the inverse tetrads 
$\lbrace e_a\,^\mu \rbrace$. The determinant of the tetrad fields is written as
$e=\det(e^a\,_\mu)$.)\par
\bigskip

The identity (\ref{1}) allows the definition of the Lagrangian density for the 
gravitational field in the TEGR, which reads (see Ref. \cite{Maluf1} for a 
review, and Refs. \cite{Combi1,Combi2} for recent analyses and criticisms)

\begin{eqnarray}
L(e) &=& -k\,e\left({1\over 4}T^{abc}T_{abc} + {1\over 2}T^{abc}T_{bac} 
- T^{a}T_{a}\right) - {1\over c}L_{M}\nonumber \\
& \equiv & -ke\Sigma^{abc}T_{abc} - {1\over c}L_{M}\,,
\label{2}
\end{eqnarray}
where $k = c^3/(16\pi G)$, $L_{M}$ represents the Lagrangian density for the 
matter fields, and $\Sigma^{abc}$ is defined by
\begin{equation}
\Sigma^{abc} = {1\over 4}\left(T^{abc} + T^{bac} - T^{cab}\right) 
+ {1\over 2}\left(\eta^{ac}T^{b} - \eta^{ab}T^{c}\right)\,.
\label{3}
\end{equation}
Thus, the Lagrangian density is geometrically equivalent to the scalar curvature
density. The variation of $ L(e)$ with respect to $e^{a\mu}$
yields the field equations 

\begin{equation}
e_{a\lambda}e_{b\mu}\partial_\nu (e\Sigma^{b\lambda \nu} )-
e (\Sigma^{b\nu}\,_aT_{b\nu\mu}-
{1\over 4}e_{a\mu}T_{bcd}\Sigma^{bcd} )={1\over {4kc}}eT_{a\mu}\,,
\label{4}
\end{equation}
where $T_{a\mu}$ is defined by 
${{\delta L_M}/ {\delta e^{a\mu}}}=eT_{a\mu} $. As expected, the field equations
are equivalent to Einstein's equations. It is possible to verify by explicit 
calculations that the equations above can be rewritten as 

\begin{equation}
\label {5}
{1\over 2}\lbrack R_{a\mu}(e) - {1\over 2}e_{a\mu}R(e)\rbrack
={1\over {4kc}}\,T_{a\mu}\,.
\end{equation}

The field equations (\ref{4}) are covariant under local Lorentz transformations
(LLT). Obviously, this property can be verified more easily from equations (\ref{5}).
The meaning of this symmetry is that the theory can be formulated in any 
reference frame, adapted to any observer in space-time. Thus, 
there are no privileged frames on which one can construct
the theory. In equations (\ref{4}), one does not need a connection to ensure the 
covariance of the equations under LLT, although in equations (\ref{5}) there appears 
the Levi-Civita connection $^0\omega_{\mu ab}(e)$. The meaning 
of LLT in a theory is precisely this: the theory can be formulated in any 
frame. Therefore, the theory is valid for all observers in space-time. This
is the relevance and main feature of the local Lorentz symmetry.
Finally, the inertial frames may be characterised by the acceleration tensor,
that yields the inertial accelerations (accelerations that are not due to the 
gravitational field) that act on a given frame in space-time.

\section{Surface terms in the action for the gravitational field}

In this section we recall some very interesting issues discussed by 
Faddeev \cite{Faddeev}, with respect to the action integral of the gravitational
field in the context of Einstein's formulation of general relativity. In 
the consideration of Hamilton's variational principle that leads to Einstein's
equations, one normally starts with the density $\sqrt{-g} R(g)$, where 
$R(g)$ is the scalar curvature constructed out of the metric tensor 
$g_{\mu\nu}$. In similarity to Faddeev's article, we will restrict the 
considerations to asymptotically flat space-times. In the limit 
$r \rightarrow \infty$, and for finite $x^0=t$ (we will now adopt $c=1$),  the 
asymptotically flat limit is characterised by 

\begin{equation}
g_{\mu\nu} = \eta_{\mu\nu}+ O({1/r})\,, \ \ \
\partial_\lambda g_{\mu\nu} = O({1/r^2})\,, \ \ \
^0\Gamma^\lambda_{\mu\nu} = O({1/r^2})\,,
\label{6}
\end{equation}
where $r^2=(x^1)^2 + (x^2)^2 + (x^3)^2$, $\eta_{\mu\nu}=(-1,+1,+1,+1)$, and
$^0\Gamma^\lambda_{\mu\nu}$ are the Christoffel symbols. The energy-momentum
tensor $T_{\mu\nu}$ for the matter fields must be of the order 
$T_{\mu\nu}=O(1/r^4)$. This condition ensures that the matter fields are 
effectively localized in a compact region of the space.

For large values of the radial coordinate $r$, the asymptotic form of the 
coordinate transformations is taken to be 

\begin{equation}
{x^\prime}^\mu=\eta^\mu(x)\,,
\label{7}
\end{equation}
where

\begin{eqnarray}
\eta^\mu(x) &=& \Lambda^\mu\,_\nu\,x^\nu + a^\mu + O(1/r)\,, \nonumber \\
\partial_\nu \eta^\mu &=& \Lambda^\mu\,_\nu + O(1/r^2)\,.
\label{8}
\end{eqnarray}
The quantities 
$\Lambda^\mu\,_\nu$ are matrices of the Lorentz transformations, and 
$a^\mu$ is an arbitrary constant vector (we are adopting Faddeev's
original notation). Faddeev assumes that these 
transformations act on the metric tensor and on the connection referred to a
fixed coordinate system. The resulting infinitesimal transformations are given
by

\begin{eqnarray}
\delta g_{\mu\nu}&=& -\partial_\mu \varepsilon^\lambda\,g_{\lambda\nu}
-\partial_\nu \varepsilon^\lambda g_{\lambda \mu}
-\varepsilon^\lambda \partial_\lambda g_{\mu\nu}\,, \nonumber \\
\delta\,(\,^0\Gamma^\lambda_{\mu\nu})&=&
-\partial_\mu \varepsilon^\sigma (^0\Gamma^\lambda_{\sigma\nu})
-\partial_\nu \varepsilon^\sigma (^0\Gamma^\lambda_{\sigma\mu})
+\partial_\sigma \varepsilon^\lambda (^0\Gamma^\sigma_{\mu\nu}) \nonumber \\
&{}&-\varepsilon^\sigma\partial_\sigma (^0\Gamma^\lambda_{\mu\nu})
-\partial_\mu \partial_\nu \varepsilon^\lambda\,,
\label{9}
\end{eqnarray}
where $\varepsilon^\lambda$ is an infinitesimal vector field that in the 
limit $r \rightarrow \infty$ has the asymptotic form given by equations (\ref{8}).

Faddeev considered the action integral constructed out of the Lagrangian density

\begin{equation}
L=\sqrt{-g}R(g)-\partial_\mu (\sqrt{-g}g^{\nu\sigma}\,^0\Gamma^\mu_{\nu\sigma}-
\sqrt{-g}g^{\mu\nu}\,^0\Gamma^\sigma_{\nu\sigma})\,,
\label{10}
\end{equation}
which differs from $ \sqrt{-g}R(g)$ by a total divergence, and argued that it is
the action constructed out of $L$,

\begin{equation}
S=\int L\;d^3x\,dt\,,
\label{11}
\end{equation}
and not the one constructed out of $\sqrt{-g}R(g)$, that is invariant under the
infinite dimensional group $G$ generated by the transformations (\ref{9}).
In view of the asymptotic behaviour given by (\ref{6}), one can verify that in 
the limit $r\rightarrow \infty$ we have

\begin{equation}
\sqrt{-g}R(g)=O(1/r^3)\,, \ \ \ \ \ L=O(1/r^4)\,.
\label{12}
\end{equation}

There is an additional essential feature of the action integral (\ref{11}).  In
the process of varying the action in order to obtain the field equations, all 
surface terms that arise in the variation of (\ref{11}), via integration by 
parts, vanish in the limit
$r \rightarrow \infty$, whereas in the variation of the action constructed out 
of $\sqrt{-g}R(g)$, several of these terms do not vanish in the same
asymptotic limit. Thus, the variation of the action constructed out of 
$\sqrt{-g}R(g)$ only 
is not well defined. Moreover, if one establishes the Hamiltonian
formulation starting from (\ref{11}), the standard ADM Hamiltonian is obtained
together with the correct surface terms that define the total ADM 
energy-momentum, i.e., one does not need to add any surface term by hand.

Similar considerations were made earlier in 1974 in the famous Lecture 
Notes by Hanson, Regge and Teitelboim \cite{Regge} on constrained Hamiltonian
systems. These authors attempted to 
write the field equations of general relativity in Hamiltonian form, i.e., the
standard Hamilton's equations in the phase space of the theory. Besides the
constraint equations, there are the evolution equations for the spatial metric
$g_{ij}$, and for the canonically conjugated momenta $\Pi^{ij}$, generated by 
the total Hamiltonian. Hanson, Regge and Teitelboim noted that one needs to add
suitable surface terms to the total Hamiltonian, so that the variation of the 
total Hamiltonian is well defined.\footnote{See pages 111-113 of Ref. 
\cite{Regge},} These surface terms are precisely the terms that yield the total
energy-momentum and angular momentum at spatial infinity. Without these
surface terms, the variation of the total Hamiltonian is not well defined, 
because of the non-vanishing of several terms that arise via integration
by parts. The improved Hamiltonian, including the
surface terms, has well defined functional derivatives.

In the context of tetrad fields $e^a\,_\mu$ and of the spin connection 
$\omega_{\mu ab}$, one also needs to add surface terms to the action integral 
in order to have well behaved functional derivatives \cite{Charap}. The 
Lagrangian density is normally considered to be $eR(e,\omega)$. This framework
is mandatory in the case of Einstein-Cartan type theories, or when one needs to
couple Dirac spinor fields to the gravitational field. 
Again, the variation of the action integral must be well defined, so that all
surface terms that arise via integration by parts vanish at spacelike infinity. 
As above, we consider the space-time to be 
asymptotically flat, and assume the asymptotic behaviour

\begin{equation}
e_{a\mu}\simeq \eta_{a\mu}+ O(1/r)\,, \ \ \ \ \omega_{\mu ab}=O(1/r^2)\,,
\label{13}
\end{equation}
in the limit $r \rightarrow \infty$. The Lagrangian density that is well defined
with respect to functional derivatives is 

\begin{eqnarray}
L(e,\omega)&=& eR(e,\omega) - \partial_\mu(e\,e^{a\mu}e^{b\nu}\omega_{\nu ab}- 
e e^{a\nu}e^{b\mu}\omega_{\nu ab}) \nonumber \\
&=&-\partial_\mu(e\,e^{a\mu}e^{b\nu})\omega_{\nu ab}+
\partial_\nu(e\,e^{a\mu}e^{b\nu})\omega_{\mu ab}\nonumber \\
&{}&+(e\,e^{a\mu}e^{b\nu})(\omega_{\mu ac}\omega_\nu\,^c\,_ b-
\omega_{\nu ac}\omega_\mu\,^c\,_ b)\,.
\label{14}
\end{eqnarray}

In the variation of $eR(e,\omega)$ alone, one finds, via integration by parts,
the term

\begin{equation}
\int dt\, d^3x \,\partial_\nu(e\,e^{a\mu}e^{b\nu}\delta \omega_{\mu ab})\neq 0\,,
\label{15}
\end{equation}
which does not vanish in general when integrated over a spatial surface at 
spacelike infinity. For a vector field $V^\alpha$ whose asymptotic behaviour in
the limit $r \rightarrow \infty$ is $V^\alpha = O(1/r^2)$, we have

\begin{eqnarray}
\int d^4 x\partial_\alpha (\sqrt{-g} V^\alpha)&=& 
\oint_{S\rightarrow \infty} dS_\alpha (\sqrt{-g} V^\alpha) \nonumber \\
&\simeq &\oint_{r=\rightarrow \infty} 
dt\, r^2 d\Omega\;\lbrack O(1/r^2)\rbrack\neq 0\,,
\label{16}
\end{eqnarray}
where $d\Omega =\sin\theta d\theta d\phi$ and, $S$ is a surface of
constant radius. The action integral constructed out 
of the Lagrangian density (\ref{14}) is not affected by this problem. In the
analysis above, it makes no difference whether we use an arbitrary spin 
connection in the Palatini variational principle, which is eventually determined
by the field equations, or the Levi-Civita connection $^0\omega_{\mu ab}(e)$.

The field equations derived from the Lagrangian (\ref{14}) are precisely 
equations (\ref{5}). Therefore, the theory determined by (\ref{14}) is covariant 
under local Lorentz transformations.

If the action integral is defined on a manifold with boundary, one may use the
ordinary Hilbert-Einstein action for the gravitational field, plus the 
Gibbons-Hawking surface term \cite{GH}, determined by the integration of the 
trace of the extrinsic curvature over the boundary.
A recent discussion on the relevance and necessity of boundary 
terms (and their variations) in the Hilbert-Einstein action is given in 
Ref. \cite{Padmanabhan} (see also the references therein). 

We note finally that the
action integral constructed out of the Lagrangian density (\ref{2}) is
not affected by the emergence of non-vanishing surface terms, in the variation 
of the action. In the analysis carried out in Ref. \cite{Maluf2}, special 
attention was paid to the need of surface terms in the action. The total
divergence in equation (\ref{1}) cancels the total divergence in equation
(\ref{14}).

\section{Teleparallel gravity with local Lorentz symmetry}

A recent article \cite{MK} summarizes a formulation of the TEGR that attempts 
to exhibit local Lorentz symmetry (see also Ref. \cite{MK2}). 
The local Lorentz symmetry is achieved by
introducing a flat space-time connection $\Omega^a\,_{b\mu}$, that corresponds
to equation (16) of Ref. \cite{MK}, and which is given by

\begin{equation}
\Omega^a\,_{b\mu}=\Lambda^a\,_c(x)\partial_\mu \Lambda_b\,^c(x)\,,
\label{17}
\end{equation}
where $\Lambda^a\,_b(x)$ are matrices of the local Lorentz group, and therefore
these quantities are space-time dependent functions. The torsion tensor, that
in Section 2 was written as 
$T_{a\mu\nu}=\partial_\mu e_{a\nu}-\partial_\nu e_{a\mu}$ (the notation for the
tetrad fields in Ref. \cite{MK} is different from ours), is now considered to be

\begin{equation}
T_{a\mu\nu}=\partial_\mu e_{a\nu}-\partial_\nu e_{a\mu}
+\Omega_{ac\mu} e^c\,_\nu-\Omega_{ac\nu} e^c\,_\mu\,.
\label{18}
\end{equation}
Although the connection (\ref{17}) is not linked to any field quantity that
has a clear transformation property, it is assumed to transform as a standard
spin connection, so that equation (\ref{18}) eventually transforms as a tensor 
under local Lorentz transformations (LLT). Except for satisfying
$\Lambda^a\,_c \Lambda^b\,_d \eta_{ab}=\eta_{cd}$, the matrices $\Lambda^a\,_ b$
are arbitrary. The authors of Ref. \cite{MK} argue that the Lagrangian density
(\ref{2}), constructed in terms of (\ref{18}), is invariant under LLT. We refer 
to Ref. \cite{MK} for additional details.

The flat spin connection (\ref{17}) is the Levi-Civita connection of the flat
space-time (see equation (112) of Ref. \cite{MK}). 
This connection is irrelevant to the dynamics of the 
tetrad field, which is the quantity that yields physical results. This fact was
already noted in Ref. \cite{Maluf2} (see equation (9) of Ref. \cite{Maluf2}.)

The counting of degrees of freedom of the flat spin connection 
$\Omega^a\,_{b\mu}$ is absolutely not clear in Ref. \cite{MK} (a recent
discussion on the degrees of freedom of the TEGR has been given in Ref.
\cite{Ferraro1}), and in the context of f(T) gravity in 
Refs. \cite{Miao,Ferraro2,Ferraro3}. This issue is 
important, because when a certain gauge is fixed, the number of degrees of 
freedom of the connection should be decreased. The vector potential $A_\mu$ in
electrodynamics, for instance, has initially 4 degrees of freedom at each 
space-time event. After fixing all gauges, the number of degrees of freedom is
reduced to 2 at each space-time event. A similar situation does not occur in the
context of Ref. \cite{MK}.

The presentation of Ref. \cite{MK} is subject to at least 5 major criticisms. 
\bigskip

\noindent 1. 
The first criticism is that a flat space-time connection (the local 
SO(3,1) group in Ref. \cite{MK} is restricted to the flat space-time) is added 
to the non-flat space-time torsion of the Weitzenb\"ock connection, as 
displayed in equation (23) of Ref. \cite{MK}. This procedure is inconsistent. A 
consistent procedure would be 
to consider a standard, ordinary affine connection subject to the 
condition of zero curvature, i.e., a regular flat connection of the local 
SO(3,1) group in the curved space-time.
This would be achieved by introducing into the Lagrangian density Lagrange
multipliers $\lambda^{ab\mu\nu}$, in order to ensure the vanishing of the 
curvature tensor constructed out of an arbitrary connection 
$\omega^a\,_{b\mu}$, i.e., $\lambda^{ab\mu\nu}\,R_{ab\mu\nu}(\omega)$. Of 
course, further consequences would result from the introduction of the Lagrange
multipliers.
\bigskip

\noindent 2. 
The second criticism is related to the variation of the action integral in the
context of Ref. \cite{MK}. According to the authors, the Lagrangian density 
that they consider, $L(e^a\,_\mu, \Omega^a\,_{b\mu})$, may be rewritten as

\begin{equation}
L(e^a\,_\mu, \Omega^a\,_{b\mu})=L(e^a\,_\mu)-
{1\over {8\pi G}}\partial_\mu(e\,\Omega^\mu)\,,
\label{19}
\end{equation}
where $\Omega^\mu =\Omega^a\,_{b\nu} e_a\,^\nu e^{b\mu}$, and $L(e^a\,_\mu)$ 
is precisely equation (\ref{2}). It is argued in 
Ref. \cite{MK} that since $\Omega^a\,_{b\mu}$ ``enters the Lagrangian as a total
derivative, the variation with respect to the spin connection vanishes 
identically''. However, the whole discussion in Section 3 was intended to show
that such variation is not trivial and non vanishing, in general. Since the 
variation of the flat connection alone is given by 
$\delta (\Omega^a\,_{b\mu})=\partial_\alpha (\Omega^a\,_{b\mu})\delta x^\alpha$, 
the integral 

\begin{equation}
\int d^4x\,\partial_\mu \lbrack e e_a\,^\nu e^{b\mu} 
\delta(\Omega^a\,_{b\nu})\rbrack=
\oint_{S\rightarrow \infty} dS_\mu 
\lbrack e e_a\,^\nu e^{b\mu} \delta(\Omega^a\,_{b\nu})\rbrack\neq 0\,,
\label{20}
\end{equation}
does not vanish, in general. In fact, if the connection $\Omega^a\,_{b\mu}$ is
constructed out of the Lorentz transformations given by the matrices in
equation (121) of Ref. \cite{MK}, for instance, then 
$\delta (\Omega^a\,_{b\mu})=O(1/r^0)$ 
everywhere in space-time. The variation above would vanish only if 
$\delta (\Omega^a\,_{b\mu})=O(1/r^3)$ in the asymptotic limit 
$r\rightarrow \infty$. Otherwise, the variation of the surface term may diverge
when integrated in the limit $r\rightarrow\infty$.

Variations of surface terms do not vanish, in general. As an example, let us 
consider the surface term that determines the total ADM mass, and which depends
on the parameter $m$ that represents the total mass of a gravitational system. 
By varying $m$, $m\rightarrow m + \delta m$, for instance, the resulting 
variation of the surface integral obviously does not vanish. In electrodynamics,
the variation of the Poynting vector, integrated over a surface of constant 
radius, is also non vanishing.
\bigskip

\noindent 3. 
Gauge theories are normally understood as constrained Hamiltonian systems, as
formulated by Dirac and summarised in Ref. \cite{Regge}. The set of first class
constraints generate the gauge transformations. This feature is connected to our
third criticism. In Ref. \cite{MK} there are no fields that would define 
first class constraints and that would yield a transformation
law for $\Omega^a\,_{b\mu}$. The gauge transformations in 
Ref. \cite{MK} are not generated by any kind of first class constraints, they 
are ``generated'' by hand. Suppose one fixes a gauge in the context of Ref. 
\cite{MK}. What would prevent the reappearance of the connection after the
gauge fixing? 

\bigskip

\noindent 4. 
According to Ref. \cite{MK}, ({\bf i}) inertial effects are represented by a 
spin connection, ({\bf ii}) there exists a class of Lorentz frames, the proper 
frames, in which there are no inertial effects, since the spin connection 
vanishes in these frames, ({\bf iii}) the flat spin connection (\ref{17}) is 
the Levi-Civita connection constructed out of the reference tetrads (see 
equation (112) of Ref. \cite{MK}). The reference tetrads are obtained by 
demanding the 
vanishing of all physical parameters of the metric tensor. These are tetrads for
the flat space-time. In Ref. \cite{MK}, the proper frames are characterised by
the vanishing of the flat spin connection.

One would expect that  for a given space-time metric tensor there would exist
a unique proper frame, i.e., a unique set of tetrad fields that is associated 
to a vanishing spin connection, where no ``spurious inertial effects'' exist.
However, it is easy to show that there may exist several distinct, physically
inequivalent tetrad fields, for a given metric tensor (specially with off 
diagonal components), that yield the same reference tetrads. Consequently, all 
these tetrads are associated to a vanishing flat spin connection, and all of 
them are supposed to be exempt of the so called spurious inertial effects. This
feature is an inconsistency, because the spurious inertial effects,
whatever it means, should vanish in the context of only one frame, the proper
frame, or in a class of frames related by global Lorentz transformations.
(The definition of inertial effects is not given in Ref. \cite{MK}.)

In order to understand this problem, let us consider the metric tensor for the
Kerr space-time in Boyer-Lindquist coordinates $(t,r,\theta,\phi)$. It is given
by

\begin{eqnarray}
ds^2&=&
-{{\psi^2}\over {\rho^2}}dt^2-{{2\chi\sin^2\theta}\over{\rho^2}}
\,d\phi\,dt
+{{\rho^2}\over {\Delta}}dr^2 \nonumber \\
&{}&+\rho^2d\theta^2+ {{\Sigma^2\sin^2\theta}\over{\rho^2}}d\phi^2\,, 
\label{21}
\end{eqnarray}
with the following definitions:

\begin{eqnarray}
\Delta&=& r^2+a^2-2mr\,, \\ \nonumber
\rho^2&=& r^2+a^2\cos^2\theta \,, \\ \nonumber
\Sigma^2&=&(r^2+a^2)^2-\Delta a^2\sin^2\theta\,, \\ \nonumber
\psi^2&=&\Delta - a^2 \sin^2\theta\,, \\ \nonumber
\chi &=&2amr\,. 
\label{22}
\end{eqnarray}

The set of tetrad fields that defines a static frame in the space-time, i.e., 
that is adapted to static observers, is given by

\begin{equation}
e_{a\mu}=\pmatrix{-A&0&0&-B\cr
0&C\sin\theta\cos\phi& \rho\cos\theta\cos\phi&-D\sin\theta\sin\phi\cr
0&C\sin\theta\sin\phi& \rho\cos\theta\sin\phi&D\sin\theta\cos\phi\cr
0&C\cos\theta&-\rho\sin\theta&0}\,,
\label{23}
\end{equation}
where

\begin{eqnarray}
A&=& {\psi \over \rho}\,,  \nonumber \\ 
B&=& {{\chi \sin^2\theta}\over {\rho \psi}}\,,  \nonumber \\ 
C&=&{\rho \over \sqrt{\Delta}}\,, \nonumber \\ 
D&=& {\Lambda \over{\rho \psi}}\,.
\label{24}
\end{eqnarray}
In the expression of $D$ we have $\Lambda =
(\psi^2\Sigma^2+\chi^2\sin^2\theta)^{1/2}\,.$

A different set of tetrad fields that satisfies Schwinger's time gauge 
condition, and which is defined from spacelike infinity up to the external 
surface of the ergosphere, reads

\begin{equation}
e_{a\mu}=\pmatrix{-A'&0&0&0\cr
B'\sin\theta\sin\phi
&C'\sin\theta\cos\phi& D'\cos\theta\cos\phi&-E'\sin\theta\sin\phi\cr
-B'\sin\theta\cos\phi
&C'\sin\theta\sin\phi& D'\cos\theta\sin\phi& E'\sin\theta\cos\phi\cr
0&C'\cos\theta&-D'\sin\theta&0}\,,
\label{25}
\end{equation}
where

\begin{eqnarray}
A'&=& {1\over \rho}\sqrt{\psi^2+{\chi^2\over \Sigma^2}\sin^2\theta}\,,
\nonumber \\
B'&=&{\chi \over {\Sigma \rho}}\,, \nonumber \\
C'&=&{\rho \over \sqrt{\Delta}}\,, \nonumber \\
D'&=&\rho\,, \nonumber \\
E'&=& {\Sigma \over \rho}\,.
\label{26}
\end{eqnarray}

These two sets of tetrad fields are adapted to observers in space-time that
have quite different 4-velocities. In equation (\ref{23}), the observers 
are static in space-time (with respect to observers at infinity). In the 
equation (\ref{25}), 
the observers have rotational motion because of the dragging effects of the 
rotating black hole. These two sets of tetrad field have quite different 
inertial properties (see the Appendix). However, both of them lead to the same
reference tetrads,

\begin{equation}
e_{a\mu}=\pmatrix{-1&0&0&0\cr
0&\sin\theta\cos\phi& r\cos\theta\cos\phi&-\sin\theta\sin\phi\cr
0&\sin\theta\sin\phi& r\cos\theta\sin\phi&\sin\theta\cos\phi\cr
0&\cos\theta&-r\sin\theta&0}\,,
\label{27}
\end{equation}
for which the flat spin connection vanishes.

Thus, from the point of view of Ref. \cite{MK}, both sets of tetrad fields above
are proper frames, and both should be free of the so called spurious inertial
effects. But it does not make sense to have two (or more) physically 
inequivalent frames, for a given space-time metric tensor, 
that are free of the same
inertial effects. (Again, disregarding the class of frames related by a global 
Lorentz transformation.) Whatever is the definition of inertial effects in Ref. 
\cite{MK}, two physically different sets of tetrad fields should exhibit
different inertial effects. From an alternative perspective, it does not make 
sense that the same flat spin connection removes the inertial effects of two 
physically inequivalent frames.
\bigskip

\noindent 5.
In addition, these frames, equations (\ref{23}) and (\ref{25}), are related by a 
local Lorentz transformation that depends on the parameters $m$ and $a$, and 
thus cannot be given by the matrices 
$\Lambda^a\,_ b$ of the flat spin connection (\ref{17}), since the latter 
matrices do not depend on physical parameters of the metric tensor, as we 
conclude from equation (112) of Ref. \cite{MK}. The flat spin connection (\ref{17})
does not provide a realisation of the full local Lorentz group SO(3,1),
since it does not generate all possible 4-rotations of the local Lorentz
group. This is one further difficulty of the formalism presented in 
Ref. \cite{MK}.

In view of all considerations above, we are led to 
conclude that the formulation of the TEGR endowed with LLT, as presented in
Ref. \cite{MK}, is inconsistent.

\section{Conclusions}

A theory is defined by the field equations, and by a set of assumptions and
interpretations of the field quantities. In this sense, the theory determined by
equation (\ref{2}) is invariant under local Lorentz transformations, as well as the 
theory determined by equation (\ref{14}). One important theoretical requirement is 
that the action integral of a theory must be well defined under functional
derivatives. One has to pay attention to surface terms that arise via 
integration by parts when varying the action. Surface terms may carry important
information about the total energy, momentum and angular momentum of the theory.

The action integral of the TEGR with local Lorentz symmetry, presented in 
Ref. \cite{MK}, is not well defined under variations of the flat spin connection
$\Omega^a\,_{b\mu}$. The variation of the action with respect to this connection
does not lead to an identically vanishing result, as the authors argue. This 
issue is a serious inconsistency. As it stands, equation (94) of Ref. \cite{MK} is
wrong. Furthermore, the LLT of Ref. \cite{MK} are not generated by first class 
constraints, as is usual in ordinary gauge theories.

The two sets of tetrads considered in the previous section, equations (\ref{23}) 
and (\ref{25}), are not related by a local Lorentz transformation generated by 
the flat spin connection (\ref{17}). These two sets of tetrads are physically
distinct. We have shown that the flat spin connection does not generate all
possible 4-rotations of the Lorentz group. Purely inertial effects of the flat
space-time cannot generate gravitational effects.

It is curious to note that in the analysis of the problem of localizability of
the gravitational energy, M\o{}ller investigated the formulation of a tetrad 
theory of gravity, and already advocated the establishment of an energy-momentum
complex that is invariant under global Lorentz transformations 
\cite{Mo1,Mo2}. In fact, he argued that not only the energy-momentum complex 
would be invariant under global (constant) Lorentz transformations, but also the
Lagrangian density itself (the particular one that he addressed) would be 
invariant under transformations with constant matrices of the Lorentz group. 
These considerations make sense, to a certain extent, because the
energy-momentum and angular momentum of the gravitational field, as well as of
any other physical system or fields, are not invariant under local Lorentz 
transformations, but covariant under global Lorentz transformations.

The physical relevance of local Lorentz transformations as a symmetry of a 
theory of gravity is that is ensures that the theory is valid in the frame of
any observer in space-time. This is an issue of consistency of the theory, that
is verified in the context of the TEGR discussed in section 2.

\appendix
\section{The acceleration tensor}

The inertial properties of any frame in space-time may be characterized by the
acceleration tensor. Here, we recall very briefly the properties of this tensor.
Let us consider that the trajectory $C$ of an observer in space-time is given by
$x^\mu(\tau)$, where $\tau$ is the proper time of the observer. The 4-velocity 
of the observer on $C$ reads $u^\mu=dx^\mu/d\tau$.
We identify the observer's velocity with the $a=(0)$ 
component of $e_a\,^\mu$: $u^\mu(\tau)=e_{(0)}\,^\mu$.
The observer's acceleration $a^\mu$ is given by the absolute 
derivative of $u^\mu$ along $C$ \cite{Hehl3},

\begin{equation}
a^\mu= {{Du^\mu}\over{d\tau}} ={{De_{(0)}\,^\mu}\over {d\tau}} =
u^\alpha \nabla_\alpha e_{(0)}\,^\mu\,, 
\label{28}
\end{equation}
where the covariant derivative
is constructed out of the Christoffel symbols. Thus, $e_a\,^\mu$
determines the velocity and acceleration along the worldline of an 
observer adapted to the frame. The set of tetrad fields for which 
$e_{(0)}\,^\mu$ describes a 
congruence of timelike curves is adapted to a class of 
observers characterized by the velocity field 
$u^\mu=e_{(0)}\,^\mu$ and by the acceleration $a^\mu$. 

We may consider not only the acceleration of observers along
trajectories whose tangent vectors are given by $e_{(0)}\,^\mu$, but
the acceleration of the whole frame along $C$. The 
acceleration of the frame is determined by the absolute derivative
of $e_a\,^\mu$ along the path $x^\mu(\tau)$. Assuming that the 
observer carries an orthonormal tetrad frame $e_a\,^\mu$, the 
acceleration of the latter along the path is given by 
\cite{Mashh2,Mashh3}

\begin{equation}
{{D e_a\,^\mu} \over {d\tau}}=\phi_a\,^b\,e_b\,^\mu\,,
\label{29}
\end{equation}
where $\phi_{ab}$ is the antisymmetric acceleration tensor. 
According to ref. \cite{Mashh2}, 
in analogy with the Faraday tensor we can identify
$\phi_{ab} \rightarrow ({\bf a}, {\bf \Omega})$, where 
${\bf a}$ is the translational acceleration ($\phi_{(0)(i)}=a_{(i)}$)
and ${\bf \Omega}$ is the angular velocity 
of the local spatial frame  with respect to a nonrotating
(Fermi-Walker transported) frame. It follows that

\begin{equation}
\phi_a\,^b= e^b\,_\mu {{D e_a\,^\mu} \over {d\tau}}=
e^b\,_\mu \,u^\lambda\nabla_\lambda e_a\,^\mu\,.
\label{30}
\end{equation}

Therefore, given any set of tetrad fields for an arbitrary 
gravitational field configuration, its geometrical interpretation
may be obtained by suitably interpreting the velocity field 
$u^\mu=\,e_{(0)}\,^\mu$ and the acceleration tensor $\phi_{ab}$.
The acceleration vector $a^\mu$ defined by equation (28) 
may be projected on a frame in order to yield

\begin{equation}
a^b= e^b\,_\mu a^\mu=e^b\,_\mu u^\alpha \nabla_\alpha
e_{(0)}\,^\mu=\phi_{(0)}\,^b\,.
\label{31}
\end{equation}
Thus, $a^\mu$ and $\phi_{(0)(i)}$ are not different 
accelerations of the frame. Along a geodesic trajectory, we have $a^\mu=0$.

It is possible to show that the acceleration tensor $\phi_{ab}$ may be
rewritten as \cite{M3,M4}

\begin{equation}
\phi_{ab}={1\over 2} \lbrack T_{(0)ab}+T_{a(0)b}-T_{b(0)a}
\rbrack\,, 
\label{32}
\end{equation}
where the torsion tensor is given by 
$T_{a\mu\nu}=\partial_\mu e_{a\nu}-\partial_\nu e_{a\mu}$.

The expression above is not invariant under local SO(3,1) 
transformations, and for this reason the values of $\phi_{ab}$ characterize 
the frame. However, equation (\ref{32}) is invariant under coordinate
transformations. We interpret $\phi_{ab}$ as the inertial (i.e., non
gravitational) accelerations along the trajectory $C$. 

The set of tetrad fields (\ref{23}) and (\ref{25}) clearly yield different
values for the acceleration tensor $\phi_{ab}$. This fact demonstrates that 
the two frames are physically inequivalent, since they are subject to different
inertial accelerations. The values of $\phi_{ab}$ for both sets of tetrads are 
very long, and for this reason we will not present them here.

\end{document}